\documentclass[twocolumn,english,showpacs,preprintnumbers,amsmath,amssymb]{revtex4}
\usepackage[latin1]{inputenc}
\usepackage{graphicx}

\makeatletter

\usepackage{natbib}
\usepackage{graphicx}
\usepackage{bm}
\usepackage{amssymb}
\usepackage{amsmath}
\def\ji{{\bm j}}
\def\dej{\delta {\bm j}}
\def\detj{\delta j}
\def\er{{\bm r}}

\def\pd{\! \cdot \!}

\makeatother
\makeatother
\makeatother
\makeatother
\makeatother
\makeatother

\usepackage{babel}
\makeatother

\newcommand{\change}[1]{{#1}}

\begin{document}

\title{Mean field theory of localization in an elongated burned fuse model}

\author{Renaud Toussaint%
\footnote{email: Renaud.Toussaint@fys.uio.no%
} and Alex Hansen%
\footnote{email: Alex.Hansen@phys.ntnu.no%
}}

\affiliation{Department of Physics, Norwegian University of Science and 
Technology, N--7491 Trondheim, Norway}
\affiliation{currently at Institut de Physique du Globe de Strasbourg, CNRS, UMR 7516,
5 rue Descartes, F-67084 Strasbourg Cedex, France}
\begin{abstract}
We propose a mean field theory for the localization of damage in a
quasistatic fuse model on a cylinder. Depending on the quenched disorder
distribution of the fuse thresholds, we show analytically that the
system can either stay in a percolation regime up to breakdown, or
start at some imposed current, to localize starting from the smallest
scale (lattice spacing), or instead go to a diffuse localization regime
where damage starts to concentrate in bands of width scaling as the
width of the system, but remains diffuse at smaller scales. Depending
on the nature of the quenched disorder on the fuse thresholds, we
derive analytically the phase diagram of the system separating these
regimes and the current levels for the onset of these possible localizations.
We compare these predictions to numerical results. 
\end{abstract}



\maketitle

\section{Introduction}

\label{intro}
To understand breakdown processes in fragile systems with elastic interactions between the elements, and disorder in the material properties, fuse networks are often studied 
\cite{HRo90,ARH85}. Such simplified models correspond to a scalar approximation of elasticity, i.e. retain the presence of long-range interactions, and such lattice models 
can be 
conveniently studied numerically, with the possibility to control a priori the probability distribution function (p.d.f.) characterizing the disorder in the rupture thresholds \cite{HRo90}.
  Fuse models allow to study the impact on breakdown processes, of parameters as the disorder in material properties, and of size effects (ratio of system size over 
lattice spacing, or over grain size for a natural system). We will present here a detailed study of an elongated fuse model, and show how three different breakdown regimes are
 accessible to it depending on the nature of the quenched disorder (q.d.) in the rupture thresholds,  and on the system size.

\begin{figure*}
\includegraphics[%
  scale=1]{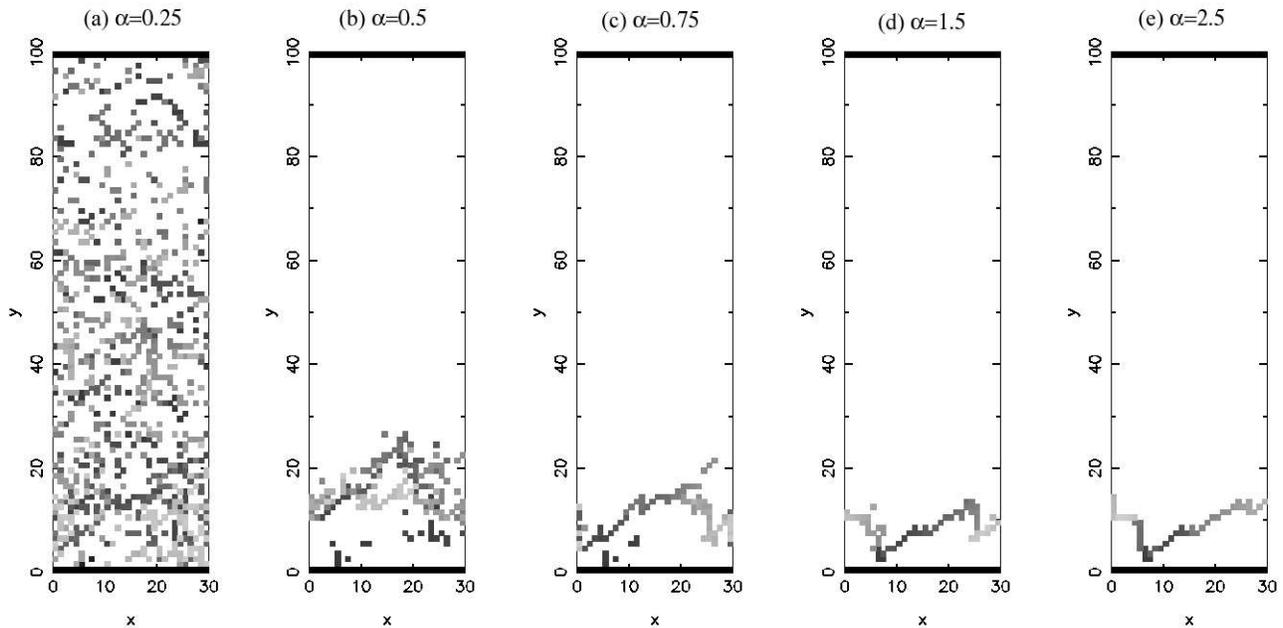}

\caption{Configuration of burned fuses in an elongated network at system breakdown, for 5 realizations with a decreasing quenched disorder 
from (a) to (e): the distribution of the fuse thresholds $t$ is of the type $p(t)\sim t^{\alpha-1}$ for $0<t<1$, with $\alpha$ indicate. In the white region underlies a nondisplayed diamond lattice of intact fuses, inclined at 45 degrees with respect to the bus bars at the top and bottom, with a lattice step $\ell=1$. Burned fuses are marked by 
gray squares, with a gray index turning from dark to light in chronological order. In cases (d) and (e), order dominates and the rupture proceeds almost always via nearest 
neighbors: this is an ordered rupture, with a {\bf total localization} of damage. In case (a), disorder is large and the breakdown process is dominated by the 
distributed location of weakest flaws: this is a {\bf percolation-like} filling, with a finite fraction of bonds to break to reach the system breakdown.  
In cases (b) and (c), rupture does not proceed via nearest neighbors, and looks diffuse at scales below $L=30 \ell$, the lateral $x-$size of the system. Broken bonds are nonetheless localized 
in a band of vertical size $w$ comparable to the horizontal width $L$ of the system: this is the {\bf diffuse localization} regime. \label{Fig:config-fuses}. }
\label{illustr-figure}
\end{figure*}

Related studies have already been performed on square fuse models \cite{HHR+91}. 
The present work extends these studies to the case of rectangular systems, with an extent $L_y$ 
importantly exceeding the dimension $L$ in the direction perpendicular to the main current flow. 
This extension will allow to show how three types of breakdown processes can emerge in it, which will be 
termed as 
an entirely localized regime, a diffuse localization regime, and a percolation-like one. We will develop an analytical mean field theory, allowing to classify which regime 
dominates the final breakdown, as function of the system size, and of the characteristics of the quneched disorder. 
The three possible regimes are illustrated in Fig.~\ref{illustr-figure}. The total localization regime corresponds 
to breakdown propagating between close or nearest neighbors. 
The percolation-like regime corresponds to systems where 
a significant fraction of the entire set of bonds have to fuse before the system becomes nonconducting. 
The diffuse localization regime corresponds to a system where the burned fuse concentrate 
in a band of size comparable with the system width, but where the damage is distributed diffusely 
inside this band, without propagating necessarily to the near neighbors of the already burned fuses.

An important motivation of this study is to characterize the scaling law between the system size, 
and the characteristic size where damage localizes in the so called ``diffuse localization'' regime. 
This scaling law has an important theoretical impact on the understanding of the origin of the geometrical characteristics of natural fracture surfaces.
Indeed, in general, the main contribution (so far) to the science of fracture by the physics community 
over the last twenty years is the discovery that brittle fracture surfaces 
are self affine \cite{MPP84}.  Self affinity implies statistical invariance 
of fracture surfaces under rescaling of length scales parallel to the average
fracture plane by a factor $\lambda$ and rescaling of the out-of-plane length
scale by a factor $\lambda^\zeta$, where $\zeta$ is the Hurst or roughness
exponent.  In 1990, based on experimental investigations of brittle aluminium 
fracture surfaces, Bouchaud {\it et al.} \cite{BLP90} proposed that the roughness
exponent has a {\it universal\/} value close to 0.8.  This value has been
reported in many later investigations, see e.g.\ 
\cite{MHHR92,SSS95,DHBC96,DNBC97}.  In Refs.\ \cite{DHBC96,DNBC97}, a small-scale
regime governed by a different roughness exponent was reported in addition to the
``usual" regime characterized by a roughness exponent of 0.8, see \cite{B97} for
a review. There have been several attempts at finding a theoretical explanation
for the universal roughness exponents, see \cite{HHR91,BB94,BBFRR02,HSc03}.

\change{Using the fuse model as paradigm for brittle fracture} \cite{ARH85,HRo90},
Hansen and Schmittbuhl \cite{HSc03} \change{have recently proposed that the roughness
exponent $\zeta$ is related to} \change{the exponent $\nu$ controlling the divergence of the
autocorrelation length of the emerging damage, $\xi$, as function of the control parameter: more explicitly, in the case of 
a burned fuse model, noting $V$ the imposed global voltage difference, and $V_c$ the voltage at complete electrical failure
,  $\xi \sim |V-V_c|^{-\nu}$. Hansen and Schmittbuhl \cite{HSc03} 
proposed the existence of a scaling relationship between these two exponents, $\zeta=2 \nu/(1+2 \nu)$ in such breakdown problems.
This relationship was numerically checked for a fuse model in two dimensions, where both the exponent $\nu$ was numerically measured, and found to be the one of percolation,
 $\nu=4/3$ \cite{Stauffer}, as well as the roughness exponent $\zeta=8/11$. 
The same reasoning for brittle fracture, 
based on the same scaling relationship, and an auto correlation divergence exponent $\nu=2$ \cite{TPr02},
leads to the roughness exponent $\zeta=4/5$ for brittle fracture, in excellent 
agreement with the experimental measurements \cite{BLP90}. Central to this theory is the
scaling law $w\sim L/\ell$ between the width of a concentrated
damage zone $w$ and the size of the system $L$, where $\ell$ is the lattice 
constant.  One of the aims of this paper is to explain the origin of this scaling law.}

The fuse model consists
of a lattice of ohmic resistors with identical conductances placed
between two bus bars, where each bond carries an electrical current
up to a threshold $t$ above which the bond burns irreversibly.
\change{Each of these local random thresholds $t$ are fixed initially and 
taken independently of each other from the p.d.f. $p(t)$, which entirely characterizes
 the uncorrelated quenched disorder
present in this system.}

 For square systems, the phase
diagram of this system was established numerically and through order statistics 
arguments \cite{HHR+91}
depending of two parameters $\alpha$ and $\beta$ characterizing
the quenched disorder distribution tails in the limit of zero or infinite
thresholds, as $p(t)\sim t^{\alpha-1}$ where $t\rightarrow0$ and
$p(t)\sim t^{-\beta-1}$ when $t\rightarrow\infty$. We will consider
here such systems in a cylindrical geometry, i.e.\ a periodic band
of finite width $L\gg\ell$, where $\ell$ is the lattice constant
placed between two bus bar at distances $L_{y}\gg L$, and derive
analytically the equivalent of this phase diagram as function of $\alpha$
and $L/\ell$, at $L_{y}/\ell\gg1$ -- we will only consider here
power-law distributions with an upper cutoff, corresponding to 
$\beta\rightarrow\infty$ in the previous terminology. With respect to this previous work, we extend the study in two ways: 
we consider elongated systems, and we in details the anisotropic aspect of the current perturbation generated by burned fuses.

In the next Section, we present the basic assumptions for and philosophy of our 
statistical analysis of the fuse model.  In Section \ref{sec:Number-density-of}
we calculate the shape of the current distribution around a region of burned-out
fuses.  We then present in Section \ref{sec:Region-of-most} the spatial
probability distribution of subsequent fuse burn-outs.  The main result of
the calculation is presented in Fig.\ \ref{Fig:phase,diag}.  Depending on the
disorder exponent $\alpha$, and on the system size, there are three possible breakdown regimes: 
(1) A percolation-like phase where no localization occurs and where a finite fraction
of the total number of fuses needs to burn out in order for the conductance of
the lattice to drop to zero in the infinite-lattice limit; (2) a  diffuse 
localization phase \change{where a damage zone develops, with a width} $w$ \change{proportional to the
width of the lattice} $L$; and (3) a complete localization phase where a single
crack evolves without damage around it.  These regimes are illustrated in Fig. 1. We do not in this paper discuss the
phase diagram with respect to the second disorder exponent $\beta$.  In
Section \ref{sec:Comparison-to-numerical}, we compare our analysis to numerical
results on the fuse model.  We summarize our
findings in Section \ref{sec:Summary}.

\section{Model Under Study and Basic Assumptions}

\label{sec:Model-under-study}

At any stage of the rupture process, we will assume that 
the local currents in the fuse model
are determined through a continuous approximation, as the solution
of the conservation of charge $\nabla\pd\ji=0$ under boundary conditions
$\ji\rightarrow j_{e}\hat{y}$ when $y\rightarrow\pm L_{y}/2$ (the
band is $L$-periodic in the $x$-direction, $(\hat{x},\hat{y})$
are the unit vectors). The current density is of the form $\ji(\er)=-c(\er)\nabla\phi(\er)$,
with a conductance $c(\er)$ equal to unity in the intact cells, and
zero in the broken ones.

After the first fuse has burnt at a certain current level $j_{e}$
at a position defined as the origin, we are interested in the average
change of external current necessary to break the next element: since
the problem is linear, for a given geometry of burnt elements, the
current flow for any other value of the external current $j'$ is
simply $(j'/j_{e})\ji(\er)$. For a given realization of the quenched
disorder $t(\er)$ (such as $t(\er)>j_{e}$ at every location), the
next fuse will burn when a first threshold is reached by the local
current, i.e.\ when the external current reaches 
\begin{equation}
j_{n}=j_{e}{\min}_{\er}\left(\frac{t(\er)}{j(\er)}\right)\;,\label{eq:def,next,ext,value}
\end{equation}
 at a position $\er_{n}$ corresponding to the realization of this
minimum. If $j_{n}>j_{e}$, the applied external current has to be
increased by a finite value for the next fuse to burn. On the contrary,
if $j_{n}\leq j_{e}$, there is an avalanche and the next fuse burns
immediately if the external current is not reduced immediately during
the first burn-out to this lower value $j_{n}$.

We are also interested in the geometric characteristics of the relative
position of the next burnt fuse with respect to the first one: over all realizations
of the quenched disorder, we define the probability distribution over
this relative position of the next burnt fuse as $\mu(\er_{n})$.
Three scenarii will be shown to happen, depending on the random mean
square distance of the next burnt fuse to the previous one, $d^{2}=\int\er_{n}^{2}\mu(\er_{n})d\er_{n}$:
(1) $d\sim+\infty$ and the process remains diffuse, resembling a
percolation process. (2) $d\sim\ell$ i.e.\ it is a function of the
lattice spacing, independent of the system width $L$. This is the
onset of a complete localization, i.e.\ the current perturbation
created by the broken cell is such that the rupture will propagate
mainly from nearest neighbor to nearest neighbor up to complete breakdown
of the system. (3) $d\sim L$, which is the onset of a regime which
we define as {}``diffuse localization:'' damage starts to concentrate
in a band of a width in the $y$ direction comparable to the system
size in the $x$ direction, $L$, but the closest neighbors of the
previously burnt cell are not significantly favorized. This is the
regime where the scaling arguments of \cite{HSc03} should apply.

If the system remains in the diffuse regime, the spatial correlations
of the damage are not significant, and we are entitled to consider
a mean field approximation to study the subsequent history of the
process: if the last fuse has burnt at a location $\er_{0}$ at a
current level $j_{e}$, the probability distribution over the location
of the next fuse burning is approximated as the probability obtained
from a situation where a single fuse has burnt at $\er_{0}$, under
the condition that all of the remaining thresholds where above $j_{e}$.

To estimate the average level of current necessary to trigger the
next fuse burning and the statistical properties of its location,
we extend the arguments of Roux and Hansen \cite{RHa90}: \change{By convention, any level of 
local current $j$ in the system will be expressed through a reduced dimensionless variable} $s=(j-j_e)/j_e$,
\change{
 ratio of the current perturbation generated by the last fuse burnt,
 over the average imposed current level.
We next define}
$n(s)\Delta s$ as the number of cells experiencing a local current
between $j_{e}(1+s)$ and $j_{e}(1+s+\Delta s)$, where $\Delta s\ll1$
is a small parameter. Defining as $\Omega(s,\Delta s)$ the region
experiencing that local current level, we have 
\begin{equation}
n(s)=\lim_{\Delta s\rightarrow0}\frac{1}{\ell^{2}\pd\Delta s}\int_{(x,y)\in\Omega(s,\Delta s)}dxdy\;.\label{eq:rel,n,Omega}\end{equation}
 The average value $m$ of the external current leading to the next
burn-out is, from Eq.\ (\ref{eq:def,next,ext,value}), the average
value of the minimum over all cells of the random variable $y=t/(1+s)$
--- Eq.\ (\ref{eq:def,next,ext,value}), 
\begin{equation}
m=\left\langle {\min}_{\left\{ s=n\Delta s,\er\in\Omega(s,\Delta s)/n\in \mathbb{Z}\right\} }\frac{t(\er)}{1+s}\right\rangle \;.\label{eq:value,of,average,min}\end{equation}
 At a given level of current perturbation $s$, we define $P(y,s)$
as the cumulative probability of the random variable $y=t/(1+s)$, given that $t>j_{e}$. \change{This last condition reflects the fact that the intact fuses 
experiencing a current $j_e(1+s)$ have survived, up to the burning point of the fuse creating the dipolar perturbation we look at.}
This is straightforwardly 
\begin{equation}
P(y,s)=\frac{P(y(1+s))-P(j_{e})}{1-P(j_{e})}\text{He}(y(1+s)-j_{e})\;,\label{eq:def,p,t,z}\end{equation}
 where $\text{He}$ is the Heaviside function, and $P$ is the cumulative
distribution of thresholds.

As shown in the Appendix, \ref{sec:statistical-lemmas.}, where we
extended some statistical results of Gumbel \cite{Gum58}, $m$ satisfies
the implicit equation \begin{equation}
\int_{s}n(s)P(m,s)ds=1\;.\label{eq:implicit,eq,for,average,min}\end{equation}
 We also show in this Appendix that $\mu(\er_{n})\,\alpha\, P(m,s(\er_{n}))$,
where $m$ is the solution of the above equation, i.e. that $n(s)P(m,s)\Delta s$
is the probability that the next bond would break in $\Omega(s,\Delta s)$.
Thus, if we find $(s_{max},\Delta s)$ such as the integral in the
above has a significant support only in $[s_{max},s_{max}+\Delta s]$
-- i.e. $\int_{s_{max}}^{s_{max}+\Delta s}n(s)P(m,s)ds=1.$, the next
break will almost certainly happen in the spatial region $\Omega(s_{max},\Delta s)$,
and the geometric properties of this spatial ensemble are representative
of the ones of the spatial distribution over all possible locations
of the next broken bond, i.e. the random mean square distance to the
next broken bond will be evaluated as \begin{equation}
d^{2}=\int_{(x,y)\in\Omega(s_{max},\Delta s)}(x^{2}+y^{2})dxdy.\label{eq:eval,practical,of,average,distance}\end{equation}

\section{Number density of cells over the level of current perturbation.}

\label{sec:Number-density-of} We will now compute the mass $n(s)\Delta s$
and shape $\Omega(s,\Delta s)$ of each region carrying a certain
value of the local current magnitude in $[j_{e}(1+s),j_{e}(1+s+\Delta s)]$.
The local current, after an unit has fused somewhere, is written as
$\ji(\er)=j_{e}\hat{y}+\dej(\er)$, with a perturbation $\dej(\er)=-\nabla\phi$
and a potential field satisfying Laplace equation $\nabla^{2}\phi=0$
under Neumann boundary conditions, $\nabla\phi=0$ when $y\rightarrow\pm L_{y}$
and $\hat{n}\,\pd\nabla\phi=j_{e}\hat{n}\,\pd\hat{\, y}$ along the
surface of the broken element (elementary lattice cell), where $\hat{n}$
is the elementary vector normal to it. Since $L\ll L_{y}$, this current
perturbation will be approximated as the one in an infinitely long
cylinder, i.e. the long range condition used will be $\nabla\phi=0$
when $y\rightarrow\pm\infty$\change{, and $x$-periodicity with a period $L$. }
We will then also use the coordinate
system where the last burnt fuse ise at the origin. 
Furthermore, from
a distance of a few lattice size and above, the shape of the lattice
cell is no more relevant, and this elementary current perturbation
is itself approximated as the solution of this problem with a spherical
fused element of diameter $\ell$: $\phi$ satisfies in circular coordinates,
$\hat{n}\pd\nabla\phi(r=\ell/2,\theta)=j_{e}\sin(\theta)$. For sufficiently
large systems $L/\ell\gg1$, this particular potential can itself
be constructed as \begin{equation}
\phi=-j_{e}\pi\ell^{2}\hat{y}\pd\nabla G/2\label{eq:const,dip,perturb}\end{equation}
 where $G$ is the solution of the Poisson equation in $L-$periodic
Boundary Conditions in the $x$ direction, satisfying $\Delta G=\delta(x,y)$
and $G(x+L,y)=G(x,y)$: indeed, along the surface of the elementary
circle of diameter $\ell$, we have $\nabla G\simeq\hat{r}/2\pi r$,
and with $\hat{r}$ the elementary radial vector, and $\theta$ the
angle between $\hat{x}$ and $\hat{r}$, $\hat{\er}\pd\nabla(\hat{y}\pd\nabla G)=\hat{\er}\pd\nabla(\sin(\theta)/2\pi r)=-2\sin(\theta)/\pi\ell^{2}$.
The complete expression of $G$ in such periodic boundary conditions
is after Morse and Feshbach \cite{MFe53}, \begin{equation}
G(x,y)=\frac{1}{4\pi}\ln\left[4\sin^{2}\left(\frac{\pi x}{L}\right)+4\sinh^{2}\left(\frac{\pi y}{L}\right)\right]\label{eq:monop,period}\end{equation}
 Eventually, at a sufficient distance from a broken cell $r\gg\ell$,
we have $\delta j\ll j_{e}$ and $j(\er)\simeq\sqrt{(j_{e}\hat{y}+\dej(\er)}\simeq j_{e}+\detj(\er)$
where $\detj(\er)=\hat{y}\pd\dej(\er)$ \change{-- which is a classical expression for the dipolar perturbation emanating from a burned fuse in such models, see e.g.} \cite{BSO02}. The magnitude of the current
perturbation is thus determined from the above Eqs. (\ref{eq:const,dip,perturb},\ref{eq:monop,period})
as \begin{eqnarray}
\detj(x,y)/j_{e} & = & \frac{\pi^{2}\ell^{2}}{2L^{2}}f(2\pi x/L,2\pi y/L)\label{eq:elem,perturb,magnit}\\
\text{with\, }f(u,v) & = & \frac{1-\cos\left(u\right)\cosh\left(v\right)}{\left(\cosh\left(v\right)-\cos\left(u\right)\right)^{2}}.\label{eq:def,f,u,v}\end{eqnarray}
 A contour map of the dimensionless current perturbation $f(u,v)$
is displayed in Figure \ref{Fig:current,perturb}. %
\begin{figure}
\includegraphics[%
  scale=0.8]{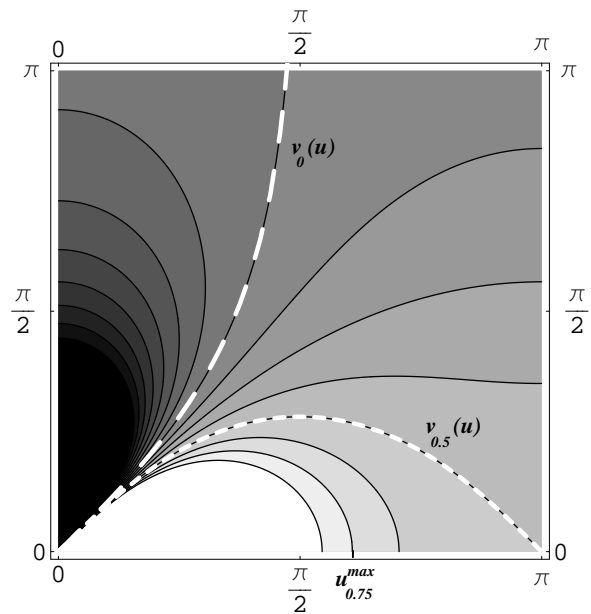}

\caption{Contour map of the elementary current perturbation due to a burnt
fuse at the origin. The long-dashed curve corresponds to a 0-perturbation,
the short-dashed one meets a saddle point at $(\pi,0)$, and corresponds
to $f(u,v)=0.5$.\label{Fig:current,perturb}}
\end{figure}

Since this perturbation $f(u,v)$ is pair in both its arguments, only
a zone $[0,\pi]^{2}$ was represented. The system is $2\pi$-periodic
in the $x$-direction. Two special contours were highlightened: $f(u,v)=0$
is the long-dashed curve. On the displayed region, points to the right
of this line experience an increased current due to the burnt fuse
at the origin, and conversely the current is screened for those to
the left of it. This zero perturbation contour correspond to $v_{0}(u)=\text{acosh}(1/\cos(u))$,
which has a support on $u\bmod[2\pi]\in[-\pi/2,\pi/2]$ and an asymptot
$v_{0}(u)\rightarrow+\infty$ when $u\rightarrow\pm\pi/2$. The other
contour goes through a saddle point of $f$ in $(u,v)=(0,\pi)$, and
corresponds to $f(u,v)=0.5$, or $v_{0.5}(u)=\text{acosh}\left(\sqrt{2-\cos^{2}u}\,\right)$.

\change{The regions $\Omega(s,\Delta s)$ that we want to characterize geometrically, which support a perturbation of current such as $s<\delta j/j_e < s+\Delta s$, correspond to the regions between two neighboring lines of the contour map on Fig.~\ref{Fig:current,perturb}. The number of the cells in such regions, defined in Eq.~(\ref{eq:rel,n,Omega}),
is shown in Appendix \ref{sec:asymptotic-values-of} to be of the form
 \begin{equation}
n(s) = \frac{2L^{4}}{\pi^{2}\ell^{4}}g(2L^{2}s/\pi^{2}\ell^{2}),\label{eq:reduction,n}
\end{equation}
where the dimensionless quantity $g$, function of its dimensionless argument, is numerically evaluated and plotted as a continuous line in Fig.~\ref{Fig:number,density,current,perturb}. The numerical evaluation is based on analytical expressions detailed in Appendix \ref{sec:asymptotic-values-of}, where are also derived the asymptotic behaviors of this function $g$ which are plotted in Fig.~\ref{Fig:number,density,current,perturb}.
}

\begin{figure}
\includegraphics[%
  scale=0.8]{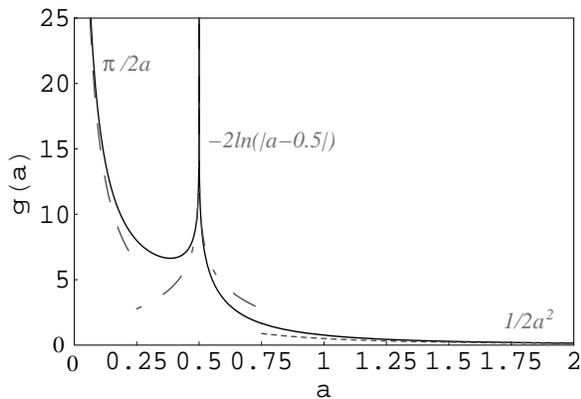}

\caption{Dimensionless number density of cells as function of the level of
current perturbation, and asymptotic forms in dashed, at infinite
distance (a=0), around the saddle point (a=1/2) and in the region
close to the burnt fuse ($a\rightarrow\infty$).\label{Fig:number,density,current,perturb}}
\end{figure}

\section{Region of Most Probable Next Event}

\label{sec:Region-of-most} 

\change{We have now derived the number of cells associated with each current level, $n(s)$, characterizing the interactions 
in this system, and need to consider some specific quenched disorder 
to determine the typical separation between two subsequent burning fuses.
With a quenched disorder distribution of power law type
$P(t)=t^{\alpha}$
on $0\leq t\leq1$, with $\alpha\geq0$, we obtain the cumulative distribution for thresholds to be below $m$, for the fuses that were still intact at current $j_e$, through }
Eq.~(\ref{eq:def,p,t,z}), as
\begin{equation}
P(m,s)=\frac{m^{\alpha}(1+s)^{\alpha}-j_{e}^{\alpha}}{1-j_{e}^{\alpha}}\text{He}(m(1+s)-j_{e}).\label{eq:specific,derived,distrib}
\end{equation}
 
\change{
We divide the space with respect to the last burned fuse in three zones, noted $\Omega_c$, $\Omega_d$ and $\Omega_f$, 
and have to solve the implicit Eq. (\ref{eq:implicit,eq,for,average,min}) to find both the most probable region out
 of these three where the next break will happen, and the most probable value of the external current $m$ at which the next fuse will burn.
This equation becomes}

\begin{equation}
(\int_{\Omega_c} + \int_{\Omega_d} + \int_{\Omega_f}) n(s) P(m,s) ds=1
\label{summary-div-integral}
\end{equation}
\change{
By definition, $\Omega_c$ is a region of fuses close to the last burned one, with the largest positive current perturbation, such as $f(u,v)>L/(2 \pi^2 \ell)$. All such fuses lie within a distance $r<\sqrt(L \ell)$ of the last burned one, where we recall that $\ell$ is the lattice step, and $L$ is the $x-$dimension of the system.}

\change{$\Omega_d$ is a region defined with moderate and finite current perturbations, where $L/(2 \pi^2 \ell)>f(u,v)>1/4$. The typical distance $r$ from the origin of the current perturbation,  over the zone $\Omega_d$, is such as $\sqrt(L \ell)<r<\pi L$.}

\change{Last, $\Omega_f$ is a region of weak to negative current perturbation, defined by the implicit equation $f(u,v)<1/4$. It will be shown that when this region 
dominates the left hand side of Eq.~(\ref{summary-div-integral}), the leading contribution to it comes from points sitting at a characteristic distance $r$ to the last burned fuse, scaling with the system size as $r\sim L_y$.}

\change{ In Appendix~\ref{integral-extreme}, we analyze in details the three terms of Eq.~(\ref{summary-div-integral}), and reformulate it as}
\begin{equation}
H_c(\lambda)+H_d(\lambda)+H_f(\lambda) = \frac{2 \varepsilon (1-j_e^\alpha)}{j_e^\alpha}
\label{eq-synthetic}
\end{equation}
\change{where $\lambda=m/j_e$, with $j_e,m$ the values of te external current at the last break and at the most probable next one,
$\varepsilon=\pi^2 \ell^2/(2L^2)$ and $H_c$, $H_d$, $H_f$ are proportional to the integrals in Eq.~(\ref{summary-div-integral}) over 
the regions $\Omega_c$, $\Omega_d$, $\Omega_f$.}

\change{We will classify the regime of the system according to the dominant term in 
the left hand side of Eq. (\ref{eq-synthetic}): If $H_f$ dominates, 
the system remains in a diffuse regime where there are no noticeable
spatial correlations in the pattern of burnt fuse. If $H_c$ dominates,
this signifys
the onset of a complete localization regime where the damage will
develop in a concentrated zone scaling as the lattice size $\ell$,
and propagate through the system, tearing it with jumps between successive events close to this smallest scale.
Last, the dominance of $H_d$ would denote the onset of a diffuse localization regime,
where the characteristic distance $d$ between the burnt fuse scales
as $L$, the system's width.}

\change{
Thus, following as the imposed current increases, which of the three terms dominates in Eq.~(\ref{eq-synthetic}), allows 
to understand when damage starts to localize, and at which spatial scale. This allows to classify, as function of the system dimensions $L_x/\ell$, $L_y/\ell$ and of the quenched disorder, characterized by $\alpha$, in which localization regime the system ends up in. }

\change{It is shown in Appendix \ref{integral-extreme} that in the early stages of the process, at small $j_e$, $H_f$ dominates the solution of Eq.~
(\ref{eq-synthetic}),
 owing to the singularity of $n(s)$ around $s\sim0$ (zero current perturbation line), and this equation reduces to}
\begin{equation} 
\lambda^{\alpha}-1 = \frac{1-j_{e}^{\alpha}}{N_{cells}j_{e}^{\alpha}}\label{eq:current,jump,percol,regime,explicit}
\end{equation}

Since the first break is typically for $j_{1}^{\alpha}=1/N_{cells}$, 
this equation predicts a second break typically at 
\begin{equation}
\lambda_{1}^{\alpha}-1=1,\label{eq:current,2nd}
\end{equation}
 i.e. the second break should happen on average at $j_{2}=\lambda_{1}j_{1}=(1+1)^{1/\alpha}j_{1}=(2/N_{cells})^{1/\alpha}$.
Since $j_{2}>j_{1}$, the process is stable and there is a finite
gap in external current to trigger the next fuse burning.
Since $H_f$ is dominated by the asymptot of zero current perturbation (noted  $h_4$ and $h_6$ in Appendix C), 
corresponding to the long dashed curve in Fig.~\ref{Fig:current,perturb}, which spans the whole $y-$ range of the system, 
the next fuse is likely to burn at a distance scaling
as $d\sim L_{y}$ from the first one, i.e. the system remains in a
diffuse regime, with no noticeable correlations between the locations
of the burnt fuses: the size of the system wins compared to the attractive
feature of the current concentration around the last burnt fuse, in
a Flory-type argument. We can then proceed with this mean-field theory
to treat the later stages of the process.

As long as $H_f$ dominates Eq.~(\ref{eq-synthetic}), Eq.(\ref{eq:current,jump,percol,regime,explicit}) remains valid, and by recurrence,
we show in Appendix \ref{integral-extreme} that the $n-th$ fuse burns on average when
the external current is such as \begin{equation}
j_{n}^{\alpha}=(n/N_{cells}):
\label{eq:external,current,percol,regime}\end{equation}
As long as this is the case, the n-th weakest bonds are the most likely to be the n first burned ones.

For threshold distributions characterized by a very large disorder, i.e.
in the limit $\alpha\rightarrow0$, it is shown in Appendix \ref{integral-extreme} that
$H_f(\lambda)$ always dominates in the solution of Eq.~(\ref{eq-synthetic}),
up to the moment where $j_{e}^{\alpha}=1/2$. In this limit, the $n-$th fuse burning corresponds to the $n-th$ weakest threshold, and this lasts until the entire system is broken due to burned fuses percolating through the system. In this case,
the process remains diffuse, in a percolation-like regime, up to the
moment where $P(j_{e})=1/2$, which corresponds
to the critical percolation threshold. This means that in this limit
of non-renormalizability of the q.d. distribution, and very large disorder, the process is
equivalent to a bond-percolation process, which was shown by Roux
et al. \cite{RHH+88}.

On contrary, for very small disorder, in the limit $\alpha \rightarrow +\infty$, we show in Appendix \ref{integral-extreme} that 
$H_c(\lambda)$ dominates, and even that the contribution of the nearest neighboring cells on the sides of the last burned fuse
 dominate the integral: thus, the next fuses to burn are the ones carrying the highest current perturbation, and from the asymptotic expression of $H_c(\lambda)$ derived in  Appendix \ref{integral-extreme}, and Eq.~(\ref{eq-synthetic}),
the level of next break is set by 
\begin{eqnarray}
\lambda &= &\frac{2^{1/\alpha}}{j_{e}s(\alpha)^{1/\alpha}}\label{eq:next,break,total,loc}
\\
\text{where\, }s(\alpha) & = & \int_{\frac{\ell}{4\pi L}}^{\frac{1}{4\pi}}\frac{(1+\gamma)^{\alpha}}{2\gamma^{2}}d\gamma . \label{eq:def,s,of,alpha}
\end{eqnarray}
 This happens in a controlled way, i.e. for $\lambda>1$
if $\alpha$ is still sufficiently small so that $s(\alpha)<2N=2/j_{e}^{\alpha}$
, or through immediate avalanches $\lambda<1$ in the opposite case.
This is the limit of no disorder, where all bonds share the same threshold,
and the concentration of current around the first broken one is the
significant parameter controlling the process in this case: the rupture
proceeds from the smallest scales, expanding through nearest neighbors
from the initial seed to tear the system apart. This corresponds to a classical rupture process, analog
to the rupture of a perfectly elastic and homogeneous material (no disorder), where the stress concentration at the tips of an initial
 default leads to the rupture of the system when the load is increased - in  stable way or not, depending on the load level --
 the situation known from one century in linear elastic fracture mechanics, treated by Griffith and Inglis \cite{Lawn}.

Between these two extreme cases, in the range of finite $\alpha$, the system can be driven to a third
regime if $H_d$ dominate in the solution of Eq. (\ref{eq-synthetic}):
correlations in the damage start to be significant, but the characteristic
distance to the preceding burnt fuses is in a range between $\sqrt{L\ell}$
and $L$, and does not scale as the lattice constant $\ell$: this is
the regime which we refer to as ``diffuse localization''.

We determine a lower value $\alpha_{m}$ of the exponent of the q.d.
distribution separating systems entirely equivalent to percolation
up to breakdown, and these leading to diffuse localization, as follows:
as long as the percolation regime holds, the value of external current,
and the size of the jumps $\lambda$ in it, are determined by Eq.
(\ref{eq:external,current,percol,regime}). This regime goes on as long
as $H_d(\lambda)$ can indeed be neglected in front of $H_f(\lambda)$.
If both terms become equal, the system transits towards the diffuse
localization regime, which is shown in Appendix \ref{integral-extreme} to
correspond to leading order in $1/N_{cells}$, to the condition
\begin{equation}
\frac{\alpha}{2}\ln(\frac{L}{\ell})=2\frac{1-j_{e}^{\alpha}}{j_{e}^{\alpha}}.
\label{eq:cond,trans,perco,diff,loc}
\end{equation}
 If this condition is not met at percolation threshold $j_{e}^{\alpha}=1/2$,
i.e. if
\begin{equation}
\alpha<\alpha_{m}=\frac{4}{\ln(L/\ell)},\label{eq:crit,alpha,percol,diffloc}
\end{equation}
 the system always remains in the percolation universality class.
If on contrary $\alpha>\alpha_{m}$, the system undergoes a transition
towards diffuse localization at a typical external current
\begin{equation}
j_{t}=1/(1+\alpha\ln(L/\ell)/4)^{1/\alpha}.\label{eq:crit,current,percol,diffloc}
\end{equation}

Similarly, we determine an upper cutoff $\alpha_{M}$ of the exponent
of the q.d. distribution, above which complete localization will prevail
about the diffuse one. By equating $H_f(\lambda)$ and $H_c(\lambda)$, with $\lambda$ evaluated from the percolation regime
expression in Eq.~(\ref{eq:current,jump,percol,regime,explicit}), we show in Appendix \ref{integral-extreme} that
this upper cutoff satisfies the implicit equation
\begin{equation}
\frac{s(\alpha_{M})-2\pi L/\ell}{\alpha_{M}}=\frac{\ln(L/\ell)}{2}.
\label{eq:implicit,eq,crit,alpha,diffloc,totloc}
\end{equation}
 From the expression of $s$, the equation above has a single solution
at $\alpha_{M}=1$. If $\alpha>1$, the system will transit towards
complete localization at a characteristic current level
\begin{equation}
j_{t}=1/(1+(s(\alpha)-2\pi L/\ell)/2)^{1/\alpha}.\label{eq:crit,current,totloc}
\end{equation}

Eventually, if $L/\ell<e^{4}\simeq54$, the above would lead to $\alpha_{M}<\alpha_{m}$,
and no diffuse localization is obtained for any value of $\alpha$.
There is instead a transition directly from the percolation regime
below $\alpha<\alpha_{d}$ to a regime leading to complete localization
for $\alpha>\alpha_{d}$, with
\begin{equation}
s(\alpha_{d})-\frac{2\pi L}{\ell}=\frac{\pi}{\varepsilon}\ln(\frac{1}{a_{m}})(\lambda_{perco}^{\alpha}-1)=2.
\label{eq:implicit,eq,crit,alpha,perco,totloc}
\end{equation}
where $\lambda_{perco}$ is estimated by Eq.~(\ref{eq:current,jump,percol,regime,explicit}).
 For $\alpha>\alpha_{d}$, the system starts a complete localization
at a current level given by Eq. (\ref{eq:crit,current,totloc}).

To summarize the above results, a phase diagram of the system, showing
the regime through which it will go to final breakdown, is shown in
Figure \ref{Fig:phase,diag}. The value of $\alpha_{d}$ was determined
numerically from Eq. (\ref{eq:implicit,eq,crit,alpha,perco,totloc}).
A visual representation of sequences of burning fuses, for small systems, in 5 points of this phase diagram, 
illustrating the three regimes, is given in Fig.~\ref{Fig:config-fuses}.
\begin{figure}
\includegraphics[%
  scale=0.8]{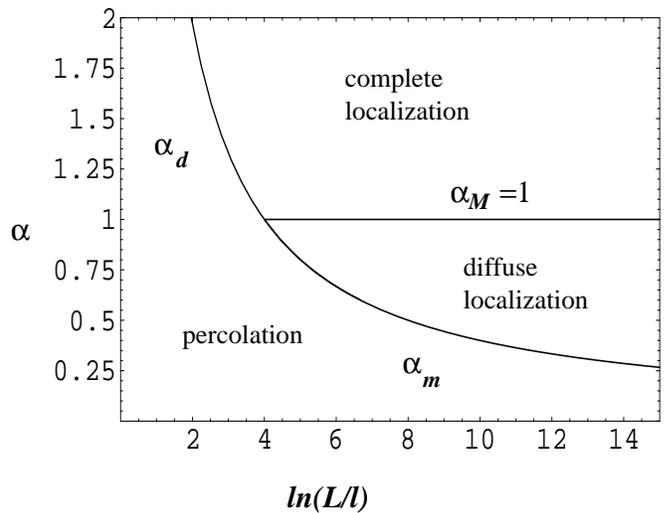}
\caption{Phase diagram of the system displaying the regime through which it
will go to macroscopic breakdown, as function of the system's width
and exponent characterizing the fuse threshold distribution\label{Fig:phase,diag}. }
\end{figure}

The above can be compared, in the limit of infinitely large systems,
to the numerical analysis carried by Hansen et al. \cite{HHR+91}:
using the notations of this paper, $\alpha=\phi_{0}$, and $1/\phi_{\infty}=0$,
and the system goes from a disorderless regime A when $\alpha>\alpha_{M}=1$
to a scaling regime B with diffuse damage and localization when $\alpha<\alpha_{M}$.
The difference between the critical exponent separating the two regimes,
which is $\alpha_{M}=1$ in the present case, and $\phi_{0}=2$ in
the models of \cite{HHR+91}, is believed to come from the elongated
character of the systems considered here ($L_{y}\gg L$).

Some remarks can be done on the succession of approximations carried to establish this phase diagram:
Most of these approximations correspond to keeping the leading order in  the inverse of the number of lattice cells in the lateral dimension, $\ell/L$,
 for systems of infinite anisotropic ratio (such as $L/L_y\rightarrow 0$). 
These approximations already take into account finite size effects, since they keep finite $\ell/L$.
 They should thus be valid as long as these numbers are $\ell \ll L$, and $L \ll L_y$.
 Evaluating the following terms corresponding  to higher orders of the parameters $\ell/L$ and  $L/L_y$ , 
in order to estimate the quality of the asymptotic expansions reduced to leading order, is beyond the scope of the present work.
In this analytical development, there is however an approximation that does not fall in this category of asymptotic expansion: 
in order to evaluate whether the system departs from the percolation like regime, as more and more fuses are burned and the system has stayed so far in this regime,
 we have considered separately the current perturbation triggered by each burned fuse. This is perfectly justified in the early stages of the process, 
where since the process is in a percolation like regime, successively burned fuses are at distances of order $L_y$ from each other, and almost do not interact. 
However, as the density of burned fuses increases, it can become finite (for sizes and q.d. where  the process always remains in a percolation like regime, 
the density reaches eventually a large fraction, in principle the percolation threshold). In this situation of high fraction of burned fuses, 
the approximation corresponding to evaluate the local current as a homogeneous background $j_e$, superimposed to the perturbation emanating from a single burned 
fuse (the last burned one), becomes of lower quality, due to the existence of multiple close burned fuses anywhere in the system. 
Overcoming this limitation at high fractions of burned fuses, would require to take into account a large number of perturbation sources. 
This task seems more suitable for a purely numerical approach: a main scope of this paper is to carry out the analytical development 
in terms of extreme statistics for a mean field theory going beyond the purely homogeneous description of damage (i.e. incorporating a 
homogeneous term plus a local perturbation). Carrying out analytically the details of the calculation while keeping the track of every 
local perturbation is beyond the scope of this work.
The result of this approximation is mainly to underestimate the number of close perturbation sources anywhere in the system as the process goes on: thus, 
this overestimates the weight of the probability to break far from the existing sources, $H_f$, and underestimates the probability to enter a 
diffuse localization or a total localization regime. Thus, the mean field theory presented here should predict properly the transition between 
''total'' and ''diffuse'' localization regime, but should over estimate the domain of the ''percolation-like'' regime: in Fig.~(\ref{Fig:phase,diag}),
 the left line should be located at smaller sizes. This approximation seems to overestimate the transition size $L/\ell$ by a finite factor not 
exceeding an order of magnitude, as will be shown in the next section.

Eventually, we note that in the diffuse localization regime, the process looks uncorrelated
at the lattice constant scale, i.e. looks like a percolation system, but
the arguments developed in this paper show that damage starts to concentrate
in a band of width scaling as the width of the system $L$. An argument
based on percolation in a gradient corresponding to the structure
of the damage concentration at the scale of the system can then be
applied to describe the breakdown process, which sustains the arguments
developed in \cite{HSc03} to explain the origin of the roughness
of the ultimate breakdown connected fronts in this regime.

Qualitatively, the phase diagram shown above supports the idea that the failure of natural macroscopic heterogeneous systems
is dominated by  either the "total localization", or the "diffuse localization" regime. Indeed, macroscopic materials
are often systems much larger
that the typical scale of the disorder, i.e. systems with a high ratio of system size over cell size, $L/ell$.
In such regime, the present work predicts that the percolation regime vanishes. More precisely, in this 
limit, the percolation regime would only subsist in the limit of non-normalizable threshold distribution, corresponding to 
$\alpha \rightarrow 0$.
So the present work predicts that the breakdown of such system is "totally localizing" at low disorder, or 
 "diffusely localizating" at larger one.
This picture is consistent with the fracture properties of natural objects:
when the fracturing solid is more homogeneous, or has only moderately disordered
toughness properties, corresponding to large values of $\alpha$, the rupture is initiated on the weakest flaws, and fracture propagates
 from nearest neighbor to nearest neighbor: this is the classical picture of linear elastic fracture mechanics of a homogeneous solid, 
described here as "total localization". The fracture of such regular object, as e.g. a crystal, leaves a flat, or close to flat, 
fracture surface, as seen in Fig.~ 1(e).
Conversely, when the toughness properties of the breaking solid are more scattered, i.e. at smaller $\alpha$, 
when the heterogeneous solid is more disordered,
the rupture proceeds according to the "diffuse localization" regime (illustrated in Fig.~1(b)): 
this corresponds to the rough post mortem fracture surfaces observed in most natural materials, 
found to be self affine with a universal roughness of 0.8.

\section{Comparison to numerical simulations}

\label{sec:Comparison-to-numerical}
We now turn to confront this theory to numerical simulations
of the fuse model. We consider rectangular models of $L \times L_{y}$
cells, with high aspect ratios $L_{y}/L$ in order to be close
to the infinitely long cylinder considered so far. The lattice constant
$\ell$ is now considered as unit length, the models considered are
periodic along the transverse $x$-direction, and the rows of nodes
at both lattice boundaries are set to two constant potential values,
with a voltage drop $\Delta U$ between both regularly increased from
$0$. \change{The rows of fuses are inclined at 45 degrees with respect to the $x$ and $y$ directions.}
Current conservation (Kirchhoff equation) is required at each
node, and the current through each fuse at location $r$ connecting
neighboring nodes with a local potential drop $\Delta V$ between
them, is $\Delta V$ if the fuse is intact, or $0$ if the fuse is
burned. This allows for each configuration of burned and intact fuses,
to obtain by solving a linear system the voltage at each node, and
the corresponding current map as $j(r)$=$C(r)\Delta U$, where $C(r)$
depends on the configuration of burned and intact fuses. The linear
inversion is performed via a conjugate gradient algorithm (Hestenes-Stiefel,
Eqs. (32) to (38) in Ref. \cite{BH88}). Initially, the
system is entirely intact and random thresholds of maximum sustainable
current $j_{t}(r)$ are picked from the quenched disorder distribution,
independently for each fuse. At any stage of the process, the location
$r$ of the next fuse to burn and the corresponding value of the external
current $\Delta U$ is obtained as $\Delta U=\min_{s}[j_{t}(s)/C(s)]=j_{t}(r)/C(r)$.

\change{An example of configurations and history of burning fuses is displayed in Fig.\ref{Fig:config-fuses}, for systems of size $L \times L_y=30\times 100$, and values of $\alpha$ between 0.25 and 2.5. The characteristic features of the three regimes are examplified in these cases.}

\begin{figure}
\includegraphics[%
  width=0.95\columnwidth,
  keepaspectratio]{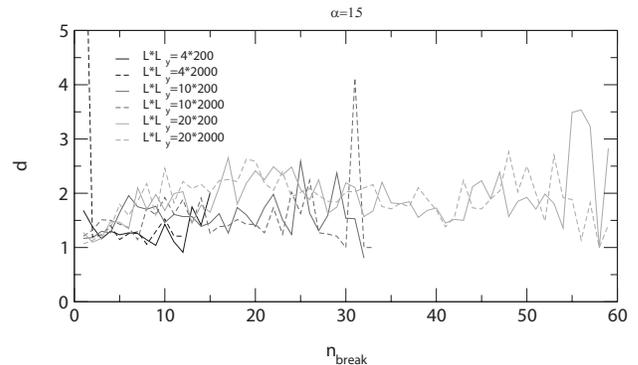}

\caption{Total localization: distance between successive burning events, as
function of their index, averaged over 50 realizations. For a quenched
disorder exponent $\alpha=1.5$, this distance remains of order unity,
irrespectively of the dimensions $L$ and $L_{y}$ of the lattice.\label{Fig:a_1_5} }
\end{figure}
For various values of $L,L_{y}$ and of the exponent $\alpha$
characterizing the quenched disorder, we look at the distance $d$
between two successive events, as function of its occurence number
in the succession of events up to total failure of the system (when
a connected line separated the upper and lower boundaries of the system).
This distance is averaged over 50 realizations. 

The characteristic situation corresponding to $\alpha>1$ is illustrated
for the case of $\alpha=1.5$ in Figure \ref{Fig:a_1_5}: the distance
between successive events is from a very early stage of order of a
few unities, irrespectively of the sizes $L$ (4, 10, 20) and
$L_{y}$ (200 and 2000) considered. This regime was referred above
as ``total localization". 

\begin{figure}
\includegraphics[%
  width=0.95\columnwidth,
  keepaspectratio]{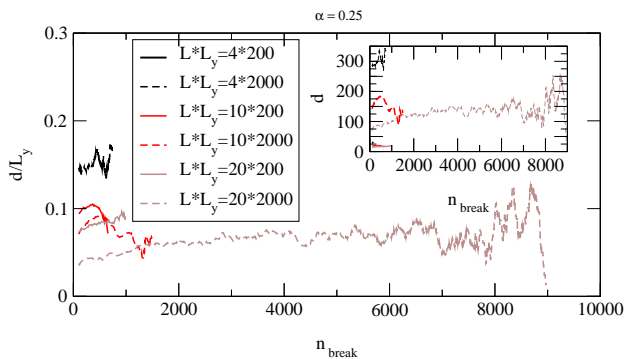}

\caption{Fuse threshold distribution corresponding to exponent $\alpha=0.25$:
percolation universality class. Distance between successive burning
fuses, divided by the lattice elongation $L_{y}$, averaged over 50
realizations and 200 successive events. Irrespectively of the lattice
dimensions, this distance is comparable to $L_{y}$ (of order $L_{y}/10$
here). Note that this corresponds depending on the lattice dimensions,
to an average distance $d$ equal to $20$ to $200$ lattice units,
as the insert shows, and equal to $1$ to $10$ times $L$\label{Fig:a_0_25}. }
\end{figure}
On contrary, for low exponents $\alpha$ corresponding to larger disorder,
the typical situation is illustrated on Figure \ref{Fig:a_0_25} by
the case of $\alpha=0.25$: Simulations have been performed using
lattice elongations $L_{y}=200$ and $L_{y}=2000$, and widths $L=4,10$
and $20$. The distance between successive events has been averaged
over 50 simulations. Even so, this quantity is still highly fluctuating,
and an additional running average over 200 successive events is performed
in order to extract the proper slowly varying average of this distance.
This high fluctuations are easily explainable: this regime is expected
to be in the universality class of percolation, where the distribution
of this distance at any stage is non negligeable for all possible
distances in the lattice. Assuming consequently that the root of the
variance of this distribution is of the same order as its average,
the central limit theorem ensures that the root of the variance of
the averaged distance over 50 realizations is still of the order 1/7
of its average, which still corresponds to a high noise to signal
ratio. The resulting average distance is plotted in figure \ref{Fig:a_0_25},
scaled by the lattice elongation $L_{y}$ in the main figure, or directly
in lattice constant units in the insert. This distance is found out to
vary slightly during the process, and shows 50\% variations between
the different probed widths $L$, but the main result is that
the average $d$ is of order $0.1L_{y}$, i.e. scales with $L_{y}$:
this is consistent with the prediction of the previous sections, that
systems of infinite elongation $L_{y}$ are isomorphic to percolation,
i.e. that the distribution of burnt-out fuses is homogeneous, irrespectively
of the configuration of the already burned fuses -- which would predict
for a very elongated system $L_{y}\rightarrow\infty$, an average
distance between successive events $d\sim\int_{0}^{L_{y}}\int_{0}^{L_{y}}dy_{1}dy_{2}|y_{1}-y_{2}|/L_{y}^{2}=L_{y}/3$.
The fact that we observe $d\sim L_{y}/10$ rather than $L_{y}/3$
can be understood as a finite size effect: less cells far away from
the last fuse burning are likely to present the minimum ratio $t/j$,
which increases the likelihood of having a next burned fuse in the
zone of significant current perturbation, closer to the last burned
fuse. 

\begin{figure}
\includegraphics[%
  width=0.95\columnwidth,
  keepaspectratio]{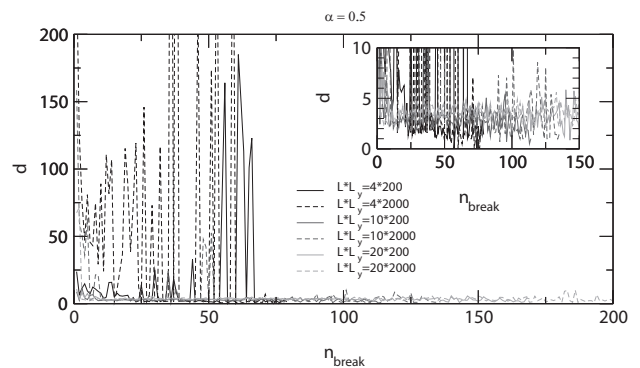}

\caption{$\alpha=0.5$: features of localization, or of percolation, depending
on $L$ and $L_{y}$: Localization (short distance between successive
events) is seen for larger $L$ and smaller $L_{y}$\label{Fig:a_0_5},
whereas episodes with distances significantly larger than $L$
are observed for short $L$ and large $L_{y}$. Data averaged
over 50 simulations.}
\end{figure}
We have also analyzed the behavior of the system for $\alpha=0.5$,
where according to Eq. ( \ref{eq:crit,alpha,percol,diffloc}), in
the limit of $L_{y}\rightarrow\infty$ one expects a percolation-like
behavior for $L<e^{4/0.5}\sim2980$, or a diffuse localization
behavior for larger system width. The average distance for $L=4,10,20$
and $L_{y}=200$ and $2000$ is displayed on figure \ref{Fig:a_0_5}
-- the insert represents the same data on a smaller scale. A qualitative
interpretation of these results can be presented as follows: focusing
first on the least elongated systems ($L_{y}=200$), the average distance
is for $L=10$ and $20$ of a few units, but the thinest systems,
$L=4$, displays a more complicated behavior: after an initial
decrease, the distance displays episodes where its magnitude is around
a few unities, alternated with episodes of order $L_{y}$. This can
be interpreted as a case on the verge between localizing or non-localizing,
i.e. as a case sitting around the line separating percolation from
localizing regime in fig. \ref{Fig:phase,diag}. The value $L=4$
is considerably smaller than the predicted $L\simeq2980$ for
infinitely elongated system. This presumably results from the underestimate in the analytical calculations, of the localizing / nonlocalizing separation
 due to the high fraction of burned fuse close to the breakdown process, and from strong finite-size
effects at finite $L_y$,  as explained in the previous Section.

 The presence of important finite-size effects is confirmed by the
fact that for more elongated systems, $L_{y}=2000$, such episodes
where the average distance significantly exceeds the width of the
system occur even more for the case $L=4$, and appear also for
the case $L=10$, while they are absent from the case $L=20$:
presumably, the boundary between localizing and percolating systems
is around $L=10$ in that case. This means at finite elongations
$L_{y}$, this boundary for any $\alpha$ corresponds to significantly
smaller $L$ than value predicted for infinitely elongated systems.
This finite size effect is observed  to
diminish for increasing elongations, as expected: the larger is the
elongation $L_{y}$, the larger is the transition width $L$.
Due to numerical costs, it seems however difficult to evaluate numerically, in the limit $L_y \rightarrow \infty$ where this finite size effects would vanish, 
the exact transition value  for the x-size separating non localizing systems and systems with diffuse localization, for example for such q.d. at $\alpha=0.5$.
From the above, at finite $L_y$, this transition size at $\alpha=0.5$ is bounded between $L=10$ (numerical observation for $L_y=2000$) and  $L \simeq 2980$ 
(theoretical upper bound from the mean field approach, Eq. (\ref{eq:crit,alpha,percol,diffloc}). 
For practical numerical purposes at moderate system sizes, we note that such systems get into localizing regimes for x-system 
sizes $L/\ell$ at one to two orders of magnitude than the previously derived upper bound, Eq. (\ref{eq:crit,alpha,percol,diffloc}).

\begin{figure}
\includegraphics[%
  width=0.95\columnwidth,
  keepaspectratio]{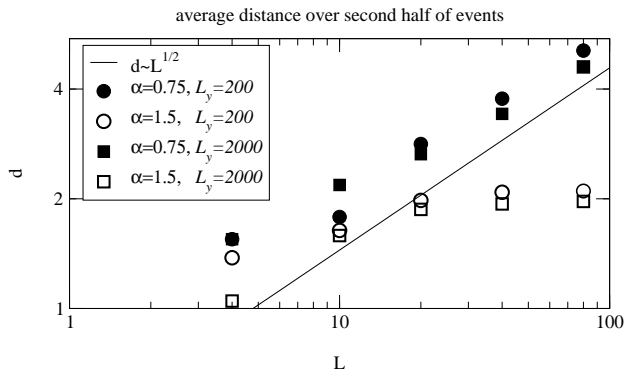}

\caption{Distinction between total and diffuse localization: at elongations
$L_{y}=200$ or $2000$, average distance $d$ between successive
events, for the second half of events, as function of the lateral
size $L$ of the system, on a bilogarithmic scale. Over the numerically
accessible range as $L$ grows, $d$ is seen to saturate for $\alpha=1.5$,
corresponding to total localization. For $\alpha=0.75$, $d/L^{1/2}$
does not vanish , corresponding to diffuse localization\label{Fig:a_0_75_and_1_5}. }
\end{figure}

Eventually, in the localizing regimes, we need to distinguish between
what was referred as diffuse or total localization in Section \ref{sec:Region-of-most}:
total localization was defined as a case where the most probable break
after departing from percolation, would happen in the zone referred
as (1), i.e. corresponding to a distance $r/\ell$ from the last burned
fuse smaller than $\sqrt{L/\ell}$ . Diffuse localization corresponds
to cases where the next event would happen preferentially in zone
(2), at moderate current perturbations, which corresponds to distances
$r/\ell$ from the last burned fuse ranging from $\sqrt{L/\ell}$
up to a few $L/\ell$. A criterion to distinguish numerically
between these two regimes is thus to look, for a fixed large elongation
$L_{y}$, whether the dependance of the average distance between successive
events over the system width is such as $d/\sqrt{L\ell}$ vanishes
at large $L$, or on contrary remains finite or diverges. On fig.
\ref{Fig:a_0_75_and_1_5} , the average distance between successive
events was evaluated over 50 simulations and over the second half
of the events before complete breakdown, which is in the localization
regime for all cases probed. This average distance seems lowly sensitive
to $L_{y}=200$ or $2000$ (20\% difference or less between both sizes),
but the scaling as function of $L$ shows that this distance saturates
rapidly for $\alpha=0.75$, while it grows approximately as $\sqrt{L}$
for $\alpha=1.5$. The extent over which this power-law corresponds
to slightly more than a decade for $L$, which is the maximum
achievable numerically since the elongation $L_{y}$ has to exceed
significantly $L$ to be in the considered framework. This result
is thus consistent with a transition from diffuse to total localization
between $\alpha=0.75$ and $\alpha=1.5$ -- the theory for $L_{y}\rightarrow\infty$
predicts this transition at $\alpha=1$. To pinpoint more accurately
the precise value of the transition exponent (between $0.75$ and
$1.5$ is not easy numerically: this would require a priori to look
at the scaling of $d(L)$ over more orders of magnitudes, for
numerous $\alpha$ inbetween, which would represent a significant
numerical cost and was not the main objective of this work, performed
mainly as a numerical check of the analytical derivations carried
out in the previous sections.

\section{Conclusion}

\label{sec:Summary}

The main results of the analytical calculation presented in this paper
are to be found in Fig.\ \ref{Fig:phase,diag}:  There are three distinct
phases of the fracture process depending on the disorder exponent $\alpha$
and on the ratio between width of the lattice $L$ and the lattice constant
$\ell$ when the lattice is a cylinder of infinite length.  The first regime
is a percolation-like regime where the distance between successive failing
fuses is completely random.  In the second regime, named ``diffuse localization",
the controlling parameter is $L/\ell$, while in the third regime, ``complete
localization", the controlling length is $\ell$.  

To summarize the numerical results of Section \ref{sec:Comparison-to-numerical},
regimes corresponding to percolation,
diffuse and total localization have been clearly identified. The transition
from diffuse to total localization is consistent with the predicted
$\alpha=1$. The transition from percolation universality class to
localizing regimes is seen to happen under increase of either the
exponent $\alpha$ of the quenched disorder power-law distribution,
or the width of the system, as predicted by the theory. Nonetheless,
the transition width for a given exponent was found significantly
smaller for the systems of finite elongation studied than for the
infinite elongation system considered analytically. This discrepancy
was observed to be lower when the elongation $L_{y}$ increased, and
presumably corresponds to important finite size effects. The transition
from diffuse to total localization is consistent with the predicted
$\alpha=1$, irrespectively of the system´s elongation.

Hence, in the diffuse localization phase, we expect a 
smooth variation of the damage profile scales $L$,
while there is still no ``strong localization'' driving the burned fuse
to merge at the lattice constant scale: region where arguments based
on percolation in gradient should apply. In this case, the shape of
the dipolar current perturbation (fig. \ref{Fig:current,perturb}),
leads to most probable relative positions $(x,y)$ of the next burned
fuse relative to the last one, for which $x$ and $y$ are of the
same order of magnitude, which is at the origin of the smooth quadratic
maximum of the damage profile as function of $y$, as observed in
\cite{HSc03}: This confirms the scaling arguments used to relate
roughness exponent and correlation length divergence exponent in such
systems.

\appendix

\section{Statistical Lemmas.}

\label{sec:statistical-lemmas.}

Consider $p$ different types of random variables $y$, characterized
by their cumulative distributions $P_{i}(y)$ and probability density
functions $p_{i}(y)=dP_{i}(y)/dy=P'_{i}(y)$, for $i=1...p$. Next,
consider an ensemble of $n_{1}$ random variables distributed according
to $p_{1}$, $n_{2}$ according to $p_{2}$,... $n_{p}$ according
to $n_{p}$. In the limit where $N=\sum_{i=1}^{p}n_{i}\gg1$, we wish
to characterize \begin{equation}
m=\left\langle {\min}_{\left\{ i=1,\dots,p,j=1,\dots,n_{i}\right\} }y_{i,j}\right\rangle \;.\label{eq:a1}\end{equation}
 The probability that some particular variable number $j$ of type
$i$, $y_{i,j}$, would be equal to $x$, while all others are larger,
is \begin{equation}
p_{i}(x)(1-P_{i}(x))^{n_{i}-1}\prod_{j\neq i}(1-P_{j}(x))^{n_{j}}\;.\label{eq:a2}\end{equation}
 The probability that any of the variables of type $i$ would be the
smallest and equal to $x$, is the above times a factor $n_{i}$.
Thus, the wanted quantity may be written as \begin{eqnarray}
 & m & =\nonumber \\
 & \int & xdx\sum_{i=1}^{p}\left[n_{i}p_{i}(x)(1-P_{i}(x))^{n_{i}-1}\prod_{j\neq i}(1-P_{j}(x))^{n_{j}}\right]\nonumber \\
 & = & \int xdx\,\,\frac{d}{dx}\left[\prod_{j=1}^{p}(1-P_{j}(x))^{n_{j}}\right]\nonumber \\
 & = & \int dx\,\,\left[\prod_{j=1}^{p}(1-P_{j}(x))^{n_{j}}\right]\;.\label{eq:a3}\end{eqnarray}

Setting $p=1$ in Eq.\ (\ref{eq:a3}), we have that \begin{equation}
m=\int dx(1-P_{1}(x))^{N}\;.\label{eq:a4}\end{equation}
 Since the function $1-P_{1}(x)$ decreases continuously from 1 to
0, for large $n$ the product $(1-P_{1}(x))^{N}$ is equal to one
for $x<x_{c}$, up to a certain cutoff $x_{c}$, above which it becomes
vanishingly small. The integral Eq.\ (\ref{eq:a4}) is then simply
equal to $x_{c}$. We determine $x_{c}$ by invoking the standard
saddle point approximation, which leads to the equation \begin{equation}
\frac{p_{1}'(x_{c})}{p_{1}(x_{c})}=(N-1)\,\,\frac{p_{1}(x_{c})}{1-P_{1}(x_{c})}\;\label{eq:a5}\end{equation}
 for $x_{c}$. By using l'H\^{o}pital's rule, $p_{1}'/p_{1}\approx p_{1}/P_{1}$,
this equation reduces to the condition $NP(x_{c})=1$. Using $m=x_{c}$,
we have \cite{Gum58}\begin{equation}
NP_{1}(m)=1\;.\label{eq:a5.5}\end{equation}
 Generalizing this result to $p>1$, we find by invoking the saddle
point approximation for Eq.\ (\ref{eq:a3}), the equation \begin{equation}
\sum_{i=1}^{p}\left(\frac{n_{i}p_{i}(m)}{1-P_{i}(m)}\right)'=\left({\sum_{j=1}^{p}\frac{n_{j}p_{j}(m)}{1-P_{j}(m)}}\right)^{2}\;.\label{eq:a6}\end{equation}
 If we now set $1-P_{j}(m)\approx1$ as $P_{j}(m)\ll1$, and use l'H\^{o}pital's
rule, \begin{equation}
\frac{\sum_{i=1}^{p}n_{i}p'_{i}(m)}{\sum_{j=1}^{p}n_{j}p_{j}(m)}\approx\frac{\sum_{i=1}^{p}n_{i}p_{i}(m)}{\sum_{j=1}^{p}n_{j}P_{j}(m)}\;,\label{eq:a6.1}\end{equation}
 Eq.\ (\ref{eq:a6}) reduces to \begin{equation}
\sum_{j=1}^{p}n_{j}P_{j}(m)=1\;,\label{eq:a8}\end{equation}
 which generalizes Eq.\ (\ref{eq:a5.5}).

In the case of an infinite number of random variables indexed by a
continuous parameter $s,$ the equivalent of Eq.\ (\ref{eq:a8})
is \begin{equation}
\int_{s}n(s)P(m,s)ds=1\;.\label{eq:a9}\end{equation}
 The probability $Q_{i}$ that the minimum variable would be of type
$i$ (for a discrete set of random variables), is from the above \begin{equation}
Q_{i}=\int dx\left[n_{i}p_{i}(x)(1-P_{i}(x))^{n_{i}-1}\prod_{j\neq i}(1-P_{j}(x))^{n_{j}}\right]\;.\label{eq:a10}\end{equation}
 The same argument shows that $(1-P_{i}(x))^{n_{i}-1}\prod_{j\neq i}(1-P_{j}(x))^{n_{j}}$
is equivalent to $1$ when $x<m$, or $0$ when $x>m$. Thus, \begin{eqnarray}
Q_{i} & = & \int_{x=-\infty}^{m}dxn_{i}p_{i}(x)\nonumber \\
 & = & n_{i}P_{i}(m)\;.\label{eq:a11}\end{eqnarray}
 For a continuous set of random variables, the probability that the
minimum variable would correspond to an index $s\in[s_{1},s_{2}]$
is then \begin{equation}
Q_{[s_{1},s_{2}]}=\int_{s_{1}}^{s_{2}}n(s)P(m,s)ds\;.\label{eq:a12}\end{equation}

\section{Density of Cells per Level of Sustained
Current}

\label{sec:asymptotic-values-of}

\change{To compute $n(s)$, we will use an explicit expression $v_{a}(u)$
of the contours of iso-perturbation, shown in Fig.~\ref{Fig:current,perturb} defined implicitly as }
\begin{equation}
f(u,v_{a}(u))=a.\label{eq:implicit,eq,for,contours}
\end{equation}
\change{Inverting this expression from Eqs. (\ref{eq:def,f,u,v},\ref{eq:implicit,eq,for,contours}),
comes the following: if $a<0$, there are two points of abcissa $u$
satisfying this. Their ordinates are }
\begin{equation}
v_{a}^{\pm}(u)=\text{acosh}\left[\frac{2a-1}{2a}\cos(u)\pm\sqrt{\frac{\cos^{2}(u)}{4a^{2}}+\frac{\sin^{2}(u)}{a}}\right]\;.\label{eq:explicit,v,a,u,a,neg}\end{equation}
 These functions are defined for \[
u\bmod[2\pi]\in[-u_{a}^{max},u_{a}^{max}]\;,\]
 with $u_{a}^{max}=\text{acos}(4a/(4a-1))$. If $0<a<0.5$, there
is a single defined function $v_{a}(u)$ satisfying Eq. (\ref{eq:implicit,eq,for,contours})
for any $u$, which is the positive alternative of the above Eq. (\ref{eq:explicit,v,a,u,a,neg}).
Last, if $a>0.5$, the expression of $v_{a}(u)$ is identical, but
once again this function has a finite support $u\bmod[2\pi]\in[-u_{a}^{max},u_{a}^{max}]$,
with $u_{a}^{max}=\text{acos}(1-1/a)$. Some examples of these auxiliary
quantities were set in Figure \ref{Fig:current,perturb}.

To compute $n(s)$, we reformulate the condition $j\in[j_{e}(1+s),j_{e}(1+s+\Delta s)]$
as $f(u,v)\in[a,a+\Delta a]$ where $a=2L^{2}s/(\pi^{2}\ell^{2})$
and $\Delta a=2L^{2}\Delta s/(\pi^{2}\ell^{2})$ from Eq. (\ref{eq:elem,perturb,magnit}),
which through a Taylor expansion of $f(u,v)$ in $v$ around $v_{a}(u)$
defines \begin{eqnarray}
\Omega(s,\Delta s)\! & \!=\! & \!\!\left\{ (u,v)/v\!\in[v_{a}(u)\!-\! w_{a}(u)\Delta a,v_{a}(u)]\right\} \label{eq:def,Omega}\\
w_{a}(u)\! & \!=\! & -1/\frac{\partial f}{\partial v}(u,v_{a}(u))\label{eq:def,Omega,thickness}\end{eqnarray}
 and thus $n(s)=\frac{1}{\Delta s\pd\ell^{2}}\int_{(x,y)\in\Omega(s,\Delta s)}dxdy$
can be expressed as
\begin{eqnarray}
n(s) & = & \frac{2L^{4}}{\pi^{2}\ell^{4}}g(2L^{2}s/\pi^{2}\ell^{2})\label{eq:reduction,n,appendix}\\
\text{where\, }g(a) & = & \int_{u=0}^{\min(u_{a}^{max},\pi)}w_{a}(u)du.\label{eq:reduced,n}\end{eqnarray}
 This number density $g(a)$ was numerically evaluated from the above
and displayed in Figure \ref{Fig:number,density,current,perturb}.
We will also derive below the asymptotic behavior of this function around the special values $a\sim0$, infinitely away from the burnt fuse, $a\rightarrow+\infty$,
around for the near neighbors of the broken fuse, and $a\sim0.5$,
around the saddle point $(u=\pi,v=0)$. As will be shown straightforwardly, 
these asymptots, displayed in Fig.~\ref{Fig:number,density,current,perturb}, are:
\begin{eqnarray}
g(a) & \sim_{a\rightarrow0} & \pi/2|a|\label{eq:asympt,far}\\
g(\delta a+1/2) & \sim_{\delta a\rightarrow0} & -2\ln(|\delta a|)\label{eq:asympt,saddle}\\
g(a) & \sim_{a\rightarrow\pm\infty} & 1/2a^{2}\;.\label{eq:asympt,neighb}
\end{eqnarray}

Indeed, from Eq.\ (\ref{eq:def,f,u,v}), \begin{eqnarray}
w_{a}(u) & = & -1/\frac{\partial f}{\partial v}(u,v_{a}(u))\label{eq:def,wa}\\
 & = & \frac{(\cosh(v_{a}(u))-\cos(u))^{3}}{\sinh(v_{a}(u))[\cos^{2}(u)+\cos(u)\cosh(v_{a}(u))-2]}\nonumber \end{eqnarray}
 where $v_{a}(u)$ is given by Eq. (\ref{eq:explicit,v,a,u,a,neg}).

Developing the above around $a\sim0^{+}$ to main order in $a$ is
a direct exercise which leads to $w_{a}(u)\sim-1/a$ when $u\in[\pi/2,\pi]$,
and $|w_{a}(u)|\ll1/a$ for $u\in[0,\pi/2]$. The integration of $w_{a}(u)$
in Eq. (\ref{eq:reduced,n}) leads then to the asymptotic value in
Eq. (\ref{eq:asympt,far}).

Conversely, around $a\sim+\infty$, $f(u,v)\sim2(u^{2}-v^{2})/(u^{2}+v^{2})^{2}=2\cos(2\theta)/r^{2}$
where . Then, defining \[
I(a)=\int_{u>0,v>0,f(u,v)<a}dudv,\]
 we have \begin{equation}
g(a)=-dI(a)/da\;,\label{eq:b2}\end{equation}
 and in polar coordinates \begin{equation}
I(a)=\int_{\theta=0}^{\pi/4}\int_{r=0}^{\sqrt{\frac{2\cos(2\theta)}{a}}}rdrd\theta=1/2a\;,\label{eq:b3}\end{equation}
 leading to the asymptotic form $g(a)\sim1/2a^{2}$, Eqa\.{ (}\ref{eq:asympt,neighb}).

Eventually, around $\delta a=a-1/2\sim0$, we reformulate the condition
$f(u,v)\in[1/2-\delta a,1/2]$ by a Taylor expansion of $f$ to second
order in $v$: $\delta f=f(u,v)-1/2=f(u,v)-f(u,v_{0.5}(u))\sim\partial_{v}f(u,v_{0.5}(u))\pd\delta v+\partial_{vv}f(u,v_{0.5}(u))\pd\delta v^{2}/2$.
The first term dominates if $\delta a<|2(\partial_{v}f)^{2}/\partial_{vv}f|$
and then $0<\delta f<\delta a$ is equivalent to $0<\delta v<-\delta a/\partial_{v}f$.
When $\delta a>2(\delta_{v}f)^{2}/\partial_{vv}f$, we will have $|\delta f|<\delta a$
if $0<\delta v<\sqrt{\delta a/\partial_{vv}f}$. From Eqs. (\ref{eq:explicit,v,a,u,a,neg},\ref{eq:def,wa}),
$v_{0.5}(u)=\sqrt{2-\cos^{2}(u)}$, and $\partial_{v}f(u,v_{0.5}(u))$
has a single zero in $u=\pi$ (saddle point), $\partial_{v}f\sim(\pi-u)/4$,
whereas $\partial_{vv}f(u,v_{0.5}(u))$ remains finite, tending towards
a finite value $\alpha$ when $u\rightarrow\pi$. Thus, we have \begin{eqnarray}
\lefteqn{\int_{u,v/f(u,v)\in[0.5-\delta a,0.5]}dudv}\label{eq:b4}\\
 & \simeq & \delta a\int_{0}^{\pi-\sqrt{\delta a/8\alpha}}\frac{-4du}{\pi-u}+O(\delta a)\nonumber \\
 & \simeq & -2\delta a\ln(\delta a)+O(\delta a)\;.\end{eqnarray}
 Differentiating this expression with respect to $\delta a$ shows
that $g(0.5-\delta a)\sim-2\ln(\delta a)$, for $\delta a\rightarrow0^{+}$.
Similar arguments lead to the same asymptotic form for a negative
$\delta a$, i.e.\ to the asymptotic form in Eq. (\ref{eq:asympt,saddle}).

\section{Spatial division of the integral characterizing extreme statistics}

\label{integral-extreme}
Here, we present the details of the subterms intervening in the implicit Eq.~(\ref{eq:implicit,eq,for,average,min}), which determine 
both the region of most probable next break with respect to the previous one, and the most probable jump in external current to reach this 
next break.

We have to solve the implicit
Eq. (\ref{eq:implicit,eq,for,average,min}), with a derived distribution
\begin{equation}
P(m,s)=\frac{m^{\alpha}(1+s)^{\alpha}-j_{e}^{\alpha}}{1-j_{e}^{\alpha}}\text{He}(m(1+s)-j_{e}).\label{eq:specific,derived,distrib,appendix}\end{equation}
coming from the assumed quenched disorder distribution, $P(t)=t^{\alpha}$
on $0\leq t\leq1$, with $\alpha\geq0$, transformed into a conditional probability of breaking for fuses having so far survived, using Eq.~(\ref{eq:def,p,t,z}).

 In the integrand of Eq. (\ref{eq:implicit,eq,for,average,min}),
$n(s)$ is expressed as function of $g(a)$ through Eq. (\ref{eq:reduction,n}),
and we approximate this last function by its asymptotic forms Eqs.
(\ref{eq:asympt,far},\ref{eq:asympt,saddle},\ref{eq:asympt,neighb})
around the singularities and in the tail of highest perturbations.

We divide the support of the integral in seven zones:

(1): in the neighboring cells carrying maximum current, such as $r/\ell<c$
where $c$ is a finite large number. The nearest neighbors on the
sides of the last broken cell carry the maximum current, $\delta j/j_{e}=f(2\pi\ell/L,0)\pd\pi^{2}\ell^{2}/2L^{2}=1/4$,
corresponding to an upper cutoff $a_{M}=L^{2}/2\pi^{2}\ell^{2}$.
Close to the origin, the current perturbation falls as $1/r^{2}$,
and this close zone is then defined by the condition $a_{M}/c^{2}<f(u,v)<a_{M}$.

(2): The cells carrying a current such as $3/4<f(u,v)<c^{2}/2\pi^{2}$.
On the $x-$axis, from Eq. (\ref{eq:def,f,u,v}), these conditions
correspond to $2L/c<x<\text{acos}(-1/3)L/2\pi\simeq L/3$. We require
that these two defined first zones have a common boundary, which by
equating the cutoffs of $f$ sets $c^{2}=L/\ell$.

(3): A zone around the saddle point, defined as $1/4<f(u,v)<3/4$

(4): A zone such as $a_{m}<f(u,v)<1/4$. This zone includes the far-range
from the last broken fuse, on which the current is slightly increased
by its presence. The lower cutoff $a_{m}$ is determined by the $y-$extent
of the system (length of the band), which was so far omitted. When
the finite aspect of $L_{y}$ is taken into account, the current perturbation
derived from boundary conditions at infinity, Eqs. (\ref{eq:elem,perturb,magnit},\ref{eq:def,f,u,v})
is still valid in boundary conditions corresponding to setting the
global current through the top and bottom plate, i.e.\  \[
\int_{x=-L/2}^{L/2}\hat{y}\pd\ji(x,\pm L_{y}/2)\, dx/L=j_{o}\;.\]
 Indeed, we find for $L_{y}\gg L$, that \begin{eqnarray*}
 & f(u,\pi L_{y}/L) & \simeq-\cos(u)/\cosh(\pi L_{y}/L)\\
 &  & \simeq-\cos(u)\exp(-\pi L_{y}/L)\;,\end{eqnarray*}
 whose integral is zero for $u\in[-\pi,\pi]$. However, when counting
the number of cells sustaining a given current, the condition $|y|<L_{y}/2$
should be added to derive $n(s)$. From the above, $n(s)$ is unmodified
when $a>a_{m}=\exp(-\pi L_{y}/L)$, or when $a<-a_{m}$, but in the
neighborhood of zero for $-a_{m}<a<a_{m}$, $n(s)$ has to be modified
according to \begin{equation}
g(a)=\frac{1}{a}\left(\text{acos}(-a/a_{m})-\pi/2\right).\label{eq:asympt,g,bounded}\end{equation}
 This function is pair, and increasing for $a\geq0$ from $g(0)=1/a_{m}$
to $g(a_{m})=\pi/2a_{m}$.

(5): A tiny zone of vanishing current perturbation, $-a_{m}<f(u,v)<a_{m}$,
for which $n(s)$ has just been determined.

(6): A far-range zone where the current is screened by the last burnt
fuse, $-1<f(u,v)<-a_{m}$.

(7): A zone of highly screened current, $-a_{M}<f(u,v)<-1$.

The zones corresponding to regions defined as $\Omega_c$, $\Omega_d$ and $\Omega_f$ in Section \ref{sec:Region-of-most}, are respectively regions (1), (2-3) and (4-7). They correspond to regions, which are with respect to the last break, either close to it, at distances comparable with the lattice step, or ``diffusively close'', at distances comparable to the width of the system $L_y$, or ``far'', at distances comparable with the system size $L_y\gg L_x$. 

Dividing here in seven subregions, will allow to use asymptotic forms of the seven corresponding subintegrals.

We will classify the regime of the system according to the zone where
most of the integral in Eq. (\ref{eq:implicit,eq,for,average,min})
is realized: We expect this zone to be either (4-6), in which case
the system remains in a diffuse regime where there are no noticeable
spatial correlations in the pattern of burnt fuse, or (1), which signifys
the onset of a complete localization regime where the damage will
develop in a concentrated zone scaling as the lattice size $\ell$,
and tear through the system starting from this smallest scale, or
(2-3), which would denote the onset of a diffuse localization regime,
where the characteristic distance $d$ between the burnt fuse scales
as $L$, the system's width.

Noting $\varepsilon=\pi^{2}\ell^{2}/2L^{2}$ and $\lambda=m/j_{e}$,
where $j_{e},m$ are the values of the external current at last break
and for the next one, Eq.\ (\ref{eq:implicit,eq,for,average,min})
is then equivalent to \begin{equation}
\sum_{i=1}^{7}h_{i}(\lambda)=\frac{2\varepsilon(1-j_{e}^{\alpha})}{j_{e}^{\alpha}}\;,\label{eq:implicit,detailed}\end{equation}
 where with \begin{equation}
p(a,\lambda)=\text{He}[\lambda(1+\varepsilon a)-1]\;,\label{eq:auxiliary,p,of,a,lambda}\end{equation}
 the $h$-functions are defined from the above zones, asymptotic behaviors
of $n(s)$ and q.d.\ distribution as \begin{eqnarray}
h_{1}(\lambda) & = & \int_{a_{M}/c^{2}}^{a_{M}}\frac{(1+\varepsilon a)^{\alpha}\lambda^{\alpha}-1}{2a^{2}}p(a,\lambda)da,\label{eq:h1}\\
h_{3}(\lambda) & = & \int_{-1/4}^{1/4}2\ln(\frac{1}{|\delta a|})[(1+\varepsilon(0.5+\delta a))^{\alpha}\lambda^{\alpha}-1]\times\nonumber \\
 &  & \qquad p(0.5+\delta a,\lambda)d\delta a,\label{eq:h3}\\
h_{4}(\lambda) & = & \int_{a_{m}}^{1/4}\frac{\pi[(1+\varepsilon a)^{\alpha}\lambda^{\alpha}-1]}{2|a|}p(a,\lambda)da,\label{eq:h4}\\
h_{5}(\lambda) & = & \int_{-a_{m}}^{a_{m}}\frac{1}{a}\left(\text{acos}(-a/a_{m})-\pi/2\right)p(a,\lambda)da\label{eq:h5}\end{eqnarray}
 and $h_{2}$ has the same integrand as $h_{1}$, but a support respectively
on $[3/4,c^{2}/2\pi^{2}]$, $h_{6}$ has the same integrand as $h_{4}$
and bounds $[-1,-a_{m}]$, $h_{7}$ has the same integrand as $h_{1}$
and has a support $[-a_{M},-1]$.

We first note that the term on the right hand-side of Eq. (\ref{eq:implicit,detailed})
is in the first stages of the process a large number, since the first
fuse burns at an average value of the external current $j_{1}$ such
as $j_{1}^{\alpha}=1/N_{cells}=\ell^{2}/LL_{y}$, so that $2\varepsilon(1-j_{1}^{\alpha})/j_{1}^{\alpha}\simeq\pi^{2}L_{y}/L\gg1$.
It becomes a number of order $\varepsilon$ when $j_{0}^{\alpha}\simeq1/2$.

We first look for solutions corresponding to an increase in external
current, i.e. we look at the behavior of the $h-$functions for $\lambda\geq1$.
The Heaviside terms can therefore be neglected for $h_{1}$, ... ,
$h_{5}$. Expanding the integrands to first order in $\varepsilon$,
and keeping the leading orders in $c=\sqrt{L/\ell}$, we obtain \begin{eqnarray}
h_{1}(\lambda) & = & \varepsilon[s(\alpha)\lambda^{\alpha}-2\pi(c^{2}-1)]\label{eq:h1,dev}\\
\text{where\, }s(\alpha) & = & \int_{\frac{1}{4\pi c^{2}}}^{\frac{1}{4\pi}}\frac{(1+\gamma)^{\alpha}}{2\gamma^{2}}d\gamma\label{eq:def,s,of,alpha,appendix}\\
h_{2}(\lambda) & = & \frac{2}{3}(\lambda^{\alpha}-1)+\frac{\alpha\varepsilon}{2}\ln(\frac{c^{2}}{2\pi})\lambda^{\alpha}\label{eq:h2,dev}\\
h_{3}(\lambda) & \simeq & (\ln(4)+1)(\lambda^{\alpha}-1)+\frac{\alpha\varepsilon}{2}\lambda^{\alpha}\label{eq:h3,dev}\\
h_{4}(\lambda) & \simeq & \frac{\pi}{2}\ln(1/a_{m})(\lambda^{\alpha}-1)+\frac{\alpha\varepsilon\pi}{8}\lambda^{\alpha}\label{eq:h4,dev}\\
h_{6}(\lambda) & \simeq & [\frac{\pi}{2}\ln(1/a_{m})(\lambda^{\alpha}-1)+\frac{\alpha\pi}{2}(1-\frac{1}{\lambda})\lambda^{\alpha}]\times\nonumber \\
 &  & \text{He}[\lambda(1+\varepsilon a_{m})-1]\label{eq:h6,dev}\end{eqnarray}
 The other $h-$functions can be shown to be negligeable in front
of these, and are not displayed here. In the early stages of the process,
the first terms of $h_{4}$ and $h_{6}$ will dominate, i.e. the singularity
of $n$ around $s\sim0$ corresponding to the furthest zone of the
last break will be preponderant, and the threshold to next break will
be set by 
\begin{eqnarray}
\pi\ln(\frac{1}{a_{min}})(\lambda^{\alpha}-1) & = & 2\varepsilon\frac{1-j_{e}^{\alpha}}{j_{e}^{\alpha}},\label{eq:current,jump,percol,regime}\\
\text{i.e.\, }\lambda^{\alpha}-1 & = & \frac{1-j_{e}^{\alpha}}{N_{cells}j_{e}^{\alpha}}\label{eq:current,jump,percol,regime,explicit,appendix}
\end{eqnarray}
which corresponds to Eq.~(\ref{eq:current,jump,percol,regime,explicit}).

 This leads to a second break happening on average when  $j_{2}=\lambda_{1}j_{1}=(1+1)^{1/\alpha}j_{1}=(2/N_{cells})^{1/\alpha}$.
As long as the process remains in a diffuse regime where burned fuse are far away from each other, at distances comparable with the $y-$ extent of the system $L_y$, we can go on with this mean field theory, and use the same arguments to evaluate the probability of finding the next ($n+1-th$) burned fuses with respect to the previously burned ones. In this situation, the fuses of interest are far from the already burned ones ($h_4$ and $h_6$ dominate the integral), and thus have thresholds above $j_e$, the external current reached so far. We can then use the same arguments, replacing the average current at first break $j_1$ by the average external current level at $n-$th break $j_n$, to obtain by reccurence that 
$j_{n}=(n/N_{cells})^{1/\alpha},$
which corresponds to Eq.~(\ref{eq:external,current,percol,regime}).
 Indeed, by definition,  $j_{n+1}=\lambda_{n}j_{n}$, and
from Eq. (\ref{eq:current,jump,percol,regime,explicit,appendix}), 
\begin{equation}
\lambda_{n}^{\alpha}-1=1/(N_{cells}j_{n}^{\alpha})=1/n.\label{eq:evol,percol}
\end{equation}
 Thus, $\lambda_{n}=[(n+1)/n]^{1/\alpha}$ and $j_{n+1}=[(n+1)/N_{cells}]^{1/\alpha}$, establishing the result by recurrence.
Note that this result is a simple consistency check of the present
theory: In this percolation regime, the concentration of the current
around the broken fuses is negligeable, and the level $j_{n+1}$ of
$(n+1)$-th break should be given by the average minimum of the thresholds
of the entire system, under the condition that all of them have survived
up to the current $j_{n}$. This formulates as $P(j_{n+1}|j_{n+1}>j_{n})=(j_{n+1}^{\alpha}-j_{n}^{\alpha})/(1-j_{n}^{\alpha})=1/N_{cells}$,
which is exactly Eq.~(\ref{eq:current,jump,percol,regime,explicit,appendix}).

A carefull analysis of the above functions shows that in the limit
y, these first terms of $h_{4}$ and $h_{6}$
always dominates: they are larger by a factor $\ln(1/a_{m})=L_{y}/L$
than the others terms proportional to $(\lambda^{\alpha}-1)$, and
all of the corrections proportional to $\varepsilon$ vanish. Thus,
in this limit, $H_f$ dominates Eq.~(\ref{eq-synthetic}), and no localization happens.
This lasts during the whole process (as long as the interactions between already burned fuses are weak enough for this mean field theory to apply), 
 and in this limit of large disorder $\alpha\rightarrow0$, the process remains diffuse, in a percolation-like regime,
 up to the moment where $j_{e}^{\alpha}=1/2$, i.e. when $P(j_{e})=1/2$, which corresponds
to the critical percolation threshold.

On contrary, for $\alpha\rightarrow+\infty$, $s(\alpha)$ is diverging
faster than all prefactors of $\lambda^{\alpha}$ in the above, and
$h_1(\lambda)$ defined in Eq.~(\ref{eq:h1,dev}), i.e.  dominates the Left Hand Side in Eq.~(\ref{eq:implicit,detailed}), i.e. $H_c(\lambda)$ dominates in Eq.~(\ref{eq-synthetic}), thus leading to a level of next break set by $h_1(\lambda)=2\varepsilon(1-j_e^\alpha)/j_e^\alpha$, i.e. Eq.(\ref{eq:next,break,total,loc}).
In this case, the next fuses to burn will be the ones carrying the highest current
perturbation, i.e. the near neighbors on the sides of the first one
(moreover, $s(\alpha)$ is dominated by the contribution between $1/4\pi$
and $1/4\pi/2^{2}$, i.e. in this limit $\alpha \rightarrow \infty$, the nearest neighbors will
burn with certainty).

In the range of finite $\alpha$, the system can be driven to a third
regime if $h_{2}$ or $h_{3}$ dominate in Eq. (\ref{eq:implicit,detailed}):
correlations in the damage start to be significant, but the characteristic
distance to the preceding burnt fuses is in a range between $\sqrt{L\ell}$
and $L$, and does not scale as the lattice constant $\ell$: this is
the regime which we refer to as ``diffuse localization''.

We have shown above that in the early stages of the process, for a finite disorder, i.e. at finite $\alpha$, the system starts in a percolation-like regime, i.e. that $H_c$ ($h_2$ and $h_3$) dominate in Eq.~(\ref{eq:implicit,detailed}). This can be the case up to the percolation transition, when $P(j_e)=1/2$,  for large enough disorder, i.e. small $\alpha$. But as the weakest bonds are broken, the process can transit to one of the two other regimes: 

either the ''total localization regime'', where rupture proceeds via jumps between successive burned fuse, whose size $d$ is close to the lattice  spacing, $d\ll \sqrt(L \ell)$ . Either the ``diffuse localization regime'', characterized by a distance $d$ between successive bonds scaling typically such as $\sqrt(L \ell) \ll d\ll L_y$ when $L\rightarrow \infty$ and $L_y \rightarrow \infty$. This transition towards the diffuse localization regime should happen when $H_d$ stops being small compared to $H_c$, while $H_f$ remains negligeable compared to these. Technically, this transition can be determined by equating $H_c(\lambda)=H_d(\lambda)$, which corresponds to leading order to $h_2(\lambda)=h_4(\lambda)+h_6(\lambda)$ (it can be shown that the contribution of the saddle point $h_{3}(\lambda)$ is always
negligeable in front of $h_{2}(\lambda)$ as long as $c^{2}=L/\ell\gg1$).
From Eqs. (\ref{eq:h2,dev},\ref{eq:h4,dev},\ref{eq:h6,dev})
corresponds to leading order in $1/N_{cells}$ to the condition
\begin{equation}
\frac{\alpha}{2}\ln(\frac{L}{\ell})=2\frac{1-j_{e}^{\alpha}}{j_{e}^{\alpha}},
\label{eq:cond,trans,perco,diff,loc,appendix}
\end{equation}
which corresponds to Eq~(\ref{eq:cond,trans,perco,diff,loc}).
As detailed in Section \ref{sec:Region-of-most}, if this condition is not met at the percolation transition when $j_e^\alpha=1/2$, the system stays in the percolation like regime, whereas if this condition is met before and $H_c$ is still negligeable, the system transits towards the diffuse localization regime before breakdown: from Eq.~(\ref{eq:cond,trans,perco,diff,loc,appendix}), this transition never happens if $\alpha<\alpha_{m}=\frac{4}{\ln(L/\ell)}$, which corresponds to Eq.~(\ref{eq:crit,alpha,percol,diffloc}).

To characterize the boundary between transition to total localization, and diffuse localization, i.e. to obtain Eq.~(\ref{eq:implicit,eq,crit,alpha,diffloc,totloc}), we have considered whether indeed $H_c(\lambda)$ is still small compared to $H_d(\lambda)$, with $\lambda$ approximated by its value at percolation threshold, i.e. fixed by Eq.~(\ref{eq:current,jump,percol,regime,explicit}) with $j_e^\alpha=1/2$. Expressing in this way the condition $h_1(\lambda)=h_2(\lambda)$, with the detailed expressions in Eqs~(\ref{eq:h1,dev}--\ref{eq:h2,dev}), together with the definitions $a_m=\exp(-\pi L_y/L)$, $\varepsilon=\pi^2 \ell^2/2L^2$ and $c^2=L/\ell$, leads to Eq.~(\ref{eq:implicit,eq,crit,alpha,diffloc,totloc}).

The estimate of the current value, Eq.~(\ref{eq:crit,current,totloc}), where the system transits towards total localization when $\alpha>1$, is obtained by equating $h_1(\lambda)=h_4(\lambda)+h_6(\lambda)$, and by noting that the last term of $h_4(\lambda)$  dominates the right hand side.

Eventually,  Eq.~(\ref{eq:implicit,eq,crit,alpha,perco,totloc}), which determines the boundary between the percolation like regime and the total localization regime in the $\alpha$-$L/\ell$ space, is obtained by equating $h_1(\lambda)=h_4(\lambda)+h_6(\lambda)$, with $\lambda$ evaluated from its value at percolation threshold,  i.e. fixed by Eq.~(\ref{eq:current,jump,percol,regime,explicit}), with $j_e^\alpha=1/2$.



\end{document}